\documentclass[preprint2]{aastex}

\newcommand{\feh}{\hbox{$ [\mathrm{Fe}/\mathrm{H}]$}}
\newcommand{\mvto}{\hbox{$ \mathrm{M_V(TO)}$}}

\shorttitle{Metal Diffusion}
\shortauthors{Chaboyer et al.}

\begin{document}

\title{Heavy Element Diffusion in Metal Poor Stars}

\author{Brian Chaboyer, W.H.\ Fenton, Jenica E.\ Nelan, D.J.\ Patnaude and
Francesca E.\ Simon }
\affil{Department of Physics and Astronomy, Dartmouth College
6127 Wilder Lab, Hanover, NH 03755}
\email{Brian.Chaboyer@Dartmouth.edu}
\email{William.H.Fenton@Dartmouth.edu}
\email{Jenica.E.Nelan@Dartmouth.edu}
\email{Daniel.J.Patnaude@Dartmouth.edu}
\email{Francesca.E.Simon@Dartmouth.edu}

\begin{abstract}
Stellar evolution models which include the effect of helium and heavy
element  diffusion have been calculated for initial iron
abundances of $\feh = -2.3,\, -2.1,\, -1.9$ and $-1.7$.  These models
were calculated for a large variety of masses and three separate
mixing lengths, $\alpha = 1.50,\, 1.75$ and $2.00$ (with $\alpha =
1.75$ being the solar calibrated mixing length). 
The change in the surface iron abundance for stars of different masses
was determined for the ages 11, 13 and 15 Gyr. 
Iron settles out of the surface convection zone on the main
sequence; this iron is dredged back up when the convection zone
deepens on the giant branch.  In all cases, the surface \feh\
abundance in the turn-off stars was at least $0.28\,$dex lower than the
surface \feh\ abundance in giant-branch stars of the same age.  
However, recent high
dispersion spectra of stars in the globular cluster NGC 6397 found
that the turn-off and giant branch stars had identical (within a few
percent) iron abundances of $\feh =-2.03$ \citep{gratton}.  These
observations prove that heavy element diffusion must be inhibited 
in the surface layers of metal-poor stars.  When diffusion is
inhibited in the outer layers of a stellar model, the predicted
temperatures of the models are similar to models evolved without
diffusion, while the predicted lifetimes are similar
to stars in which diffusion is not inhibited.  Isochrones constructed
from the models in which diffusion is inhibited fall half-way between
isochrones without diffusion, and isochrones with full diffusion.  As
a result, absolute globular cluster ages which are based upon the absolute
magnitude of the turn-off are  4\% larger than ages inferred from full
diffusion isochrones, and 4\% smaller than ages inferred from no diffusion
isochrones.  
\end{abstract}


\section{Introduction}
Atomic diffusion, whereby heavier elements settle relative to hydrogen
in a star is a physical process which  occurs in stars.
Helioseismology provides clear evidence that diffusion is occurring in
the Sun \citep[e.g.\ ][]{jcd,basu2}
As a result, a large number of studies have been made of the effect of
diffusion on the evolution of the Sun and other stars \citep[e.g.\ ][]
{nord,Michaud,michaud86,con,chab94,wind,salaris,sw2001}.  Models and
isochrones appropriate for globular cluster stars have shown that the
inclusion of diffusion reduces the main sequence life time of the
stars and reduces their effective temperatures. 
The absolute age inferred for the oldest globular clusters is reduced by 
$\sim 7\%$ when diffusion is included in the stellar models
\citep{diff1}.   

Diffusion leads to changes in the surface abundances of elements.  In
particular, the abundances of elements heavier than hydrogen will
decrease over time.  The characteristic time scale for the depletion
of an element out of the surface of a star is given by \citep{Michaud}
\begin{equation}
\tau \simeq K\frac{M_{CZ}}{MT_{CZ}^{3/2}}
\label{eqtau}
\end{equation}
where $M$ is the total mass of the star, $M_{CZ}$ is the mass of the
surface convection zone, $T_{CZ}$ is the temperature at the base of
the convection zone and $K$ is a constant, which can vary for
different elements.  Since $M_{CZ}$ can be a strong function of the
total mass of the star, the diffusion models predict that the surface
abundances vary for stars of different masses.

Observations of Li in metal-poor stars have shown the existence of a
plateau, where the abundance is constant over a wide range of
effective temperatures and metallicities \citep[e.g.\
][]{spite,thor,ryan}.  These observations suggest that diffusion is
inhibited near the surface of metal-poor stars \citep[e.g.\
][]{Michaud,con,chab94}.  However, \cite{sw2001} have shown that the
Li  abundance observations can be reproduced by models which
include fully efficient diffusion.  This follows from their analysis
which includes the effects of observational errors, uncertainties in
the Li abundance determinations and in the effective temperature
scale, and the size of the observed samples of stars.  Li is one of
three elements produced during the big bang, and so its primordial
abundance is of considerable importance in constraining big bang
nucleosynthesis.  \cite{sw2001} found that, due to diffusion, their
predicted primordial Li abundance was a factor of two higher than that
observed in the metal-poor plateau stars.

The inclusion of diffusion in stellar models has important
consequences for the estimated age of the oldest globular clusters
and the inferred primordial Li abundance.  Thus, it is important to
determine if diffusion is actually occurring in metal-poor stars.  The
recent observations of iron abundances in turn-off and giant branch
stars in the metal-poor globular cluster NGC 6397 by \cite{gratton}
provide an excellent test for the occurrence of diffusion in metal-poor
stars. \cite{gratton} found that the turn-off and giant branch stars 
had identical iron abundances (within a few percent).  
Stars near the turn-off can
have convective masses which are  orders of magnitude smaller than
those on the giant branch. Thus, one would have
expect that if diffusion does occur near the surface of metal-poor
stars, it would lead to a marked difference in the surface abundance
of iron between the main sequence turn-off stars, and those in the
giant branch.  These effects should be particularly noticeable in NGC
6397 because it is a relatively metal-poor cluster with 
$\feh = -2.02\pm 0.04$ \cite{gratton}.  The convective
envelope masses on the main sequence become smaller as one goes to
more metal-poor stars; as a result, the  diffusion time scale
(equation \ref{eqtau}) is shorter for metal-poor stars.

The use of iron abundances in globular clusters to study diffusion 
has several advantages over using Li observations in field halo
stars.  First, all the stars in NGC 6397 were  born with the same
initial composition and at the same time
\footnote{  The spectroscopic observations of 
\protect{\cite{gratton}} and  \protect{\cite{castil}}  
have found no star to star
abundance variations among the giant branch stars.  
Photometric studies \protect{\citep{Alcaino}} show a single turn-off
which indicates that all of the stars in NGC 6397 have the same age.}.
Li abundances are usually determined for field halo stars
which have different compositions, and may have different ages.
Second, the interpretation of the Li observations is complicated by
the fact that Li is destroyed at temperatures $T \ga 2.5\times
10^6\,$K, implying that mixing or turbulence within a star 
can lead to a change in the surface abundance of Li.  In
contrast, mixing or turbulence can only change the surface abundance
of iron if diffusion is already operating;  otherwise the abundance
of iron is the same everywhere in the star and mixing cannot change
the surface abundance of iron.  Finally, iron abundance measurements
are typically determined from a large number of lines, minimizing the
observational errors.  Li abundances are determined from a
single line, and the Li abundance measurements are quite sensitive to
the adopted effective temperature scale.

Section \ref{secmodel}  of this paper  discusses the stellar models
and isochrones used in this study. This includes fits of the
isochrones to the observed color magnitude digram of NGC 6397.  The
effects of metal diffusion on the surface abundance of iron are
presented in \S \ref{secmetal}, where it is concluded that diffusion
must be inhibited in the surface layers of metal-poor stars.  The
impact of inhibiting the diffusion on the derived ages of globular
clusters is explored in \S \ref{secage}.  Finally, \S \ref{secsumm}
contains a summary of the principal results in this paper.

\section{Stellar Models and Isochrones \label{secmodel}}
The stellar evolution models were calculated using Chaboyer's stellar
evolution code.  The basic input physics used in the code is described
in \cite{cgl}, with the exception of the atomic diffusion
coefficients.  These calculations use the \cite{thoul} diffusion
coefficients for helium and heavy elements.  The variation of the
abundances of H, He and Fe are followed, all other heavy elements are
assumed to diffuse at the same speed as fully-ionized iron.  The
differences in the diffusion velocities among the heavy elements are
small for the low mass stars ($M \le 0.9\,\mathrm{M}_\sun$) considered
here \citep{turcotte}, and our procedure does not introduce any
significant error into the calculations.  The local changes in the
metal abundance are taken into account in the opacity computation by
interpolating among tables with different heavy element mass fractions
($Z$).

Note that the \cite{thoul} diffusion coefficients do not include the
effects of radiative acceleration.  Studies of the effects of radiative
acceleration on solar metallicity F-stars have been performed by
\cite{richer} and \cite{turcotte}.  These studies have shown that radiative
acceleration is predicted to be important for solar metallicity stars
with masses greater than approximately $1.2\,M_\sun$.  The importance
of radiative acceleration depends on the opacity (which is a function
of temperature and density) and the radiative flux. In general the
physical conditions near the base of the surface convection zone in a 
$M = 0.80\,M_\sun$ metal-poor star are similar to those in a $M =
1.2\,M_\sun$ solar metallicity star.  As a result, it is unlikely that
radiative acceleration is important in the stars which we will model
in this paper.  Furthermore, if radiative acceleration was important,
the surface abundances of the elements would be effected, leading to
large abundance anomalies \citep{turcotte}.  Such abundance anomalies
are not observed.

A calibrated solar model was calculated by adjusting the initial mixing
length, $Z$, and the helium abundance in the model until a
$1.0\,\mathrm{M}_\sun$ model yielded the correct solar radius,
luminosity and a surface $Z/X = 0.0245$ \citep{grevesse} at the solar age
(assumed to be 4.6 Gyr).  The calibrated solar model had a mixing
length of $\alpha = 1.75$; an initial heavy element mass fraction of
$Z = 0.02$ and an initial helium abundance of $Y = 0.275$. At 4.6 Gyr,
the model had a surface $Z = 0.0179$ and $Y = 0.249$. This value for
the present day value of $Y$ is in good agreement with helioseismic
helium-abundance determinations which find $Y = 0.24 - 0.25$
\citep{basu,Richard}.  The base of the convection zone is located at
$R = 0.716\,\mathrm{R}_\sun$ in the model, which is a similar to the
depth inferred from helioseismology of $R = 0.713\,\mathrm{R}_\sun$
\citep{basu}.

The metal poor stellar models were calculated for four different
initial Fe abundances: $\feh = -2.3,\, -2.1,\, -1.9$ and $-1.7$.  The
$\alpha$-capture elements were assumed to be uniformly enhanced by
$0.2\,$dex, which is similar to the observed value in NGC 6397.
\cite{gratton} found $[\mathrm{O/Fe}] = +0.21\pm 0.05$ in their high
dispersion study of NGC 6397.  As $\alpha$-element enhanced low
temperature opacities were not available to us, a scaled solar
composition was used in the calculations.  The effect of
$\alpha$-element enhancement was taken into account by changing the
relationship between $Z$ and \feh\ \citep{Salaris}.  \cite{Salaris}
showed that most of the evolutionary properties of metal poor stars
with enhanced $\alpha$-capture elements could be reproduced by models
with scaled solar compositions if the total heavy element mass
fraction $Z$ was identical in the two sets of models.  By calculating
models over a relatively large range in \feh\, we can explore what
effect possible mis-matches between the observed abundances and those
used in our models have on predictions.  

The helium abundance in the models was determined assuming a
primordial helium abundance of $Y_p = 0.245$ and $\Delta Y/\Delta Z =
1.5$.  The primordial helium abundance was chosen to be in agreement
with \cite{Burles} who base their determinations upon big bang 
nucleosynthesis and deuterium abundance measurements in quasars.  The
value of  $\Delta Y/\Delta Z$ was determined using $Y_p$ and the solar
values of initial $Y$ and $Z$ from the models.

In order to investigate how uncertainties in the treatment of
convection affect the results, models were calculated for three
different values of the mixing length, $\alpha = 1.50,\, 1.75$ and
$2.00$ (recall that $\alpha = 1.75$ is the solar calibrated mixing
length).  Models with masses ranging from $M = 0.5 - 0.9 \,\mathrm{M}_\sun$
were evolved to  ages of 11, 13 and 15 Gyr.  The mass spacing ranged
from $0.01 \,\mathrm{M}_\sun$ to $0.001 \,\mathrm{M}_\sun$.  In total,
over 4000 stellar evolution runs were calculated.
The age range of 11 -- 15 Gyr was chosen, as it encompasses the most
recent estimate for the absolute age of old, metal-poor globular
clusters, $13.2\pm 1.5\,$Gyr \citep{chaboyer}.  

\begin{figure}[t]
\plotone{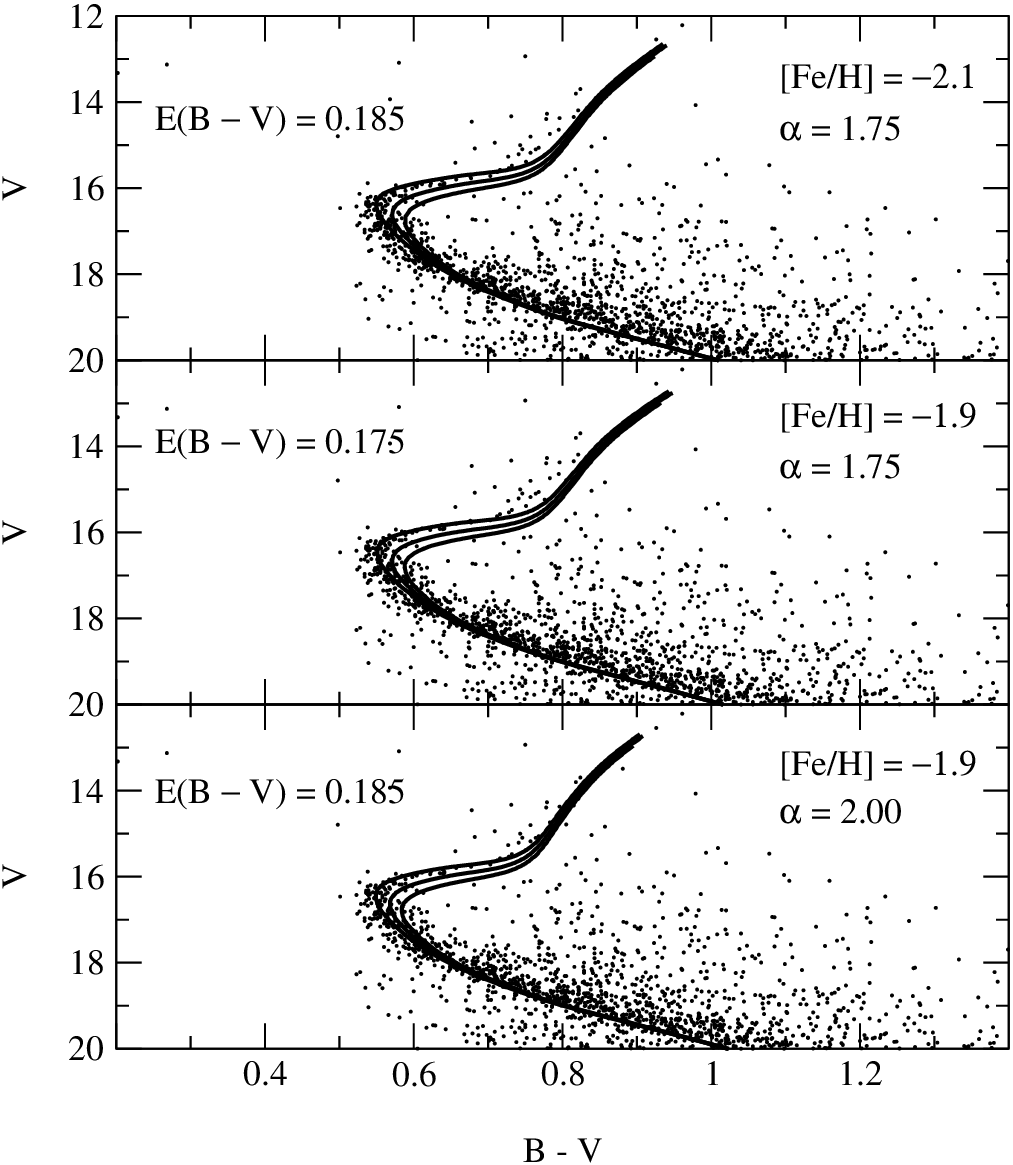}
\caption{Isochrone fits to the NGC 6397 photometry of
\protect\cite{Alcaino} for different values of \feh\ and the mixing
length $\alpha$.  All panels show 11 (bluest turn-off), 13 and 15
 Gyr
isochrones.  A distance modulus $\mathrm{(m - M)_V} = 12.60$ was used
in all of the fits.  \label{isofit}}
\end{figure}

As a test of the models, isochrones were calculated using the stellar
models which included the effects of diffusion and fit to the
NGC 6397 color-magnitude observations of \cite{Alcaino}. The isochrones were
constructed in manner described by \cite{cgl}. In performing the
isochrone fits, the distance modulus and reddening were varied until
the best fit to the unevolved main sequence was obtained.  The
reddening was constrained by the study of \cite{Twarog} who obtained
$uvby\mathrm{H}\beta$ photometry of NGC 6397 and found $\mathrm{E(B -
V)} = 0.179\pm 0.003\,$mag.  In fitting the theoretical isochrones to
the observed data, the reddening was allowed to vary by $\pm
2\,\sigma$ from the value determined by \cite{Twarog}.  The distance
scale to globular clusters is still a matter of considerable debate
\citep[e.g.\ ][]{chaboyer} and, as a result, the distance modulus was
allowed to vary by $\pm 0.15\,$mag.  Some samples of the isochrone
fits are shown in Figure \ref{isofit}.  In general, an age of 11 Gyr
was found for NGC 6397.  The best fitting isochrones had temperatures
at the turn-off ranging from $\mathrm{T_{eff}} = 6460$ to  $6630\,$K.
Based upon the H$_\alpha$ profiles in their high dispersion spectra, 
\cite{gratton} estimated the
temperature of the turn-off stars to be $\mathrm{T_{eff}} = 6576\pm
90\,$K, which is in good agreement with the models.  

\section{Metal Diffusion Results \label{secmetal}}

\begin{figure}[t]
\plotone{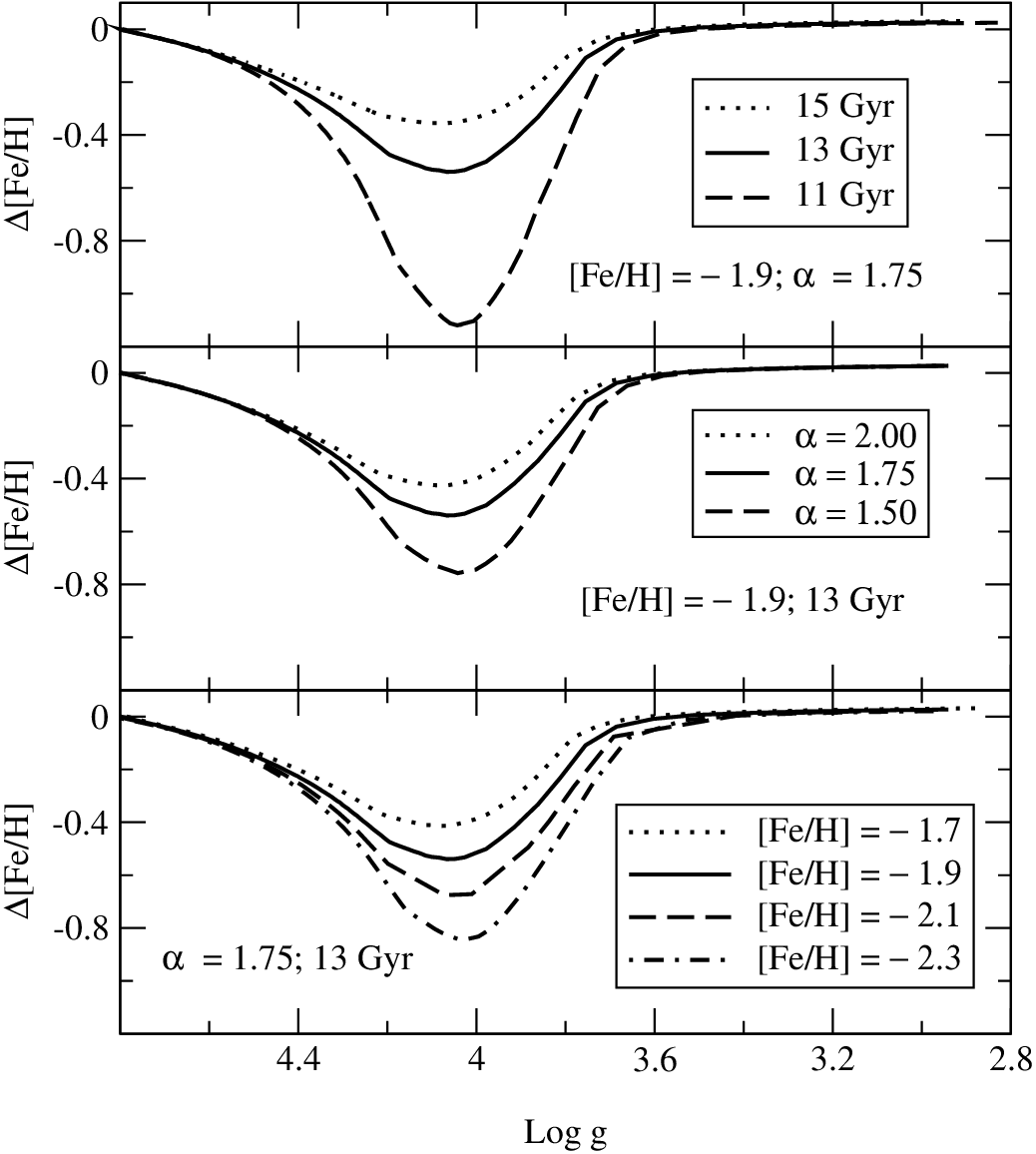}
\caption{The change in the surface \feh\ as a function of $\log g$ for
different ages (top panel),
mixing lengths (middle panel) and values of initial \feh\ (bottom
panel). The solid line is the same
in all of the  panels (initial $\feh\ = -1.9;\, \alpha = 1.75;\, 13\,$Gyr).
Main sequence stars are on
the left of this diagram, while red giant branch stars are on the
right.  \label{logg}}
\end{figure}

\begin{figure}[t]
\plotone{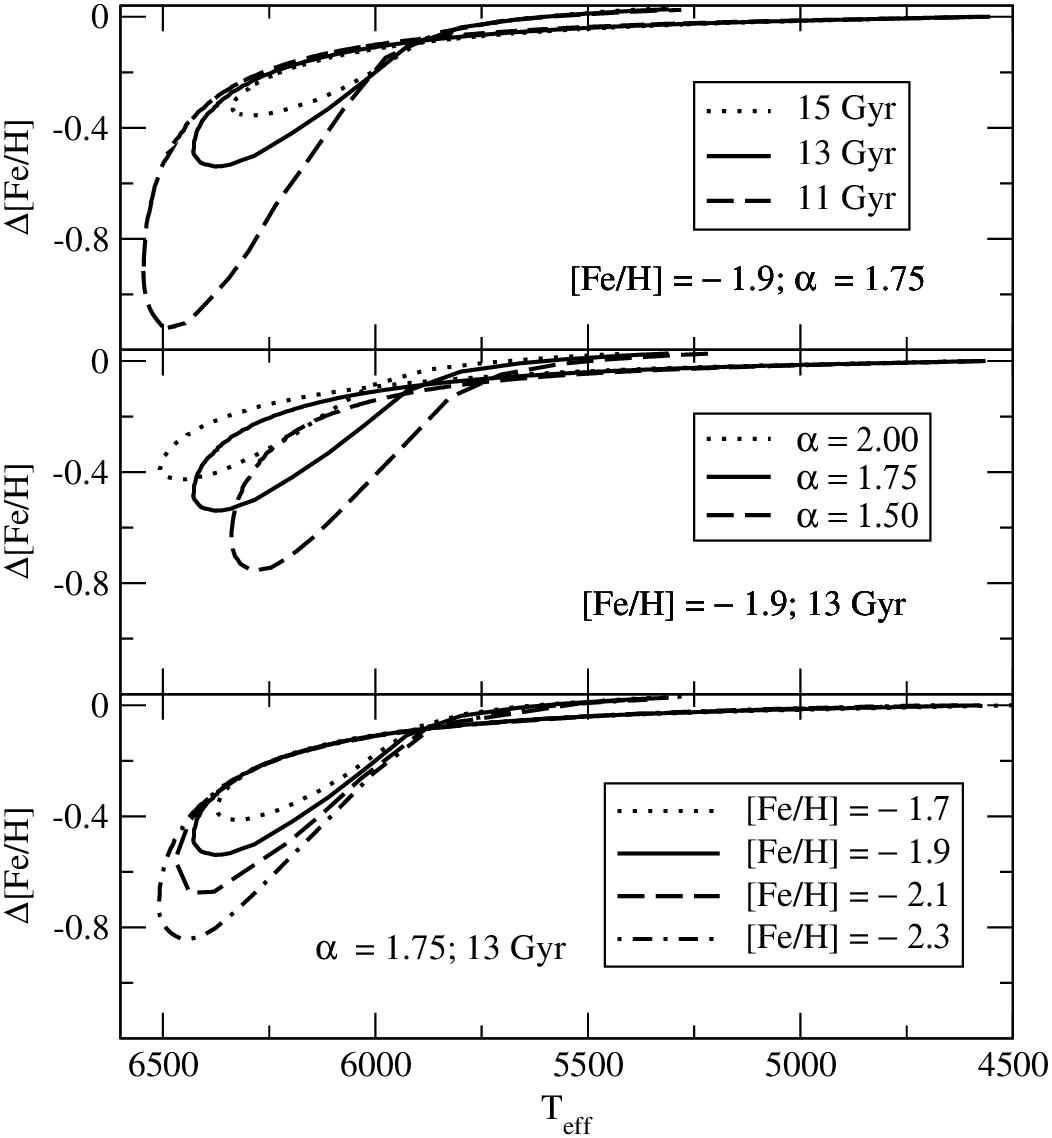}
\caption{The change in the surface \feh\ as a function of effective
temperature  for
different ages (top panel),
mixing lengths (middle panel) and values of initial \feh\ (bottom
panel). The solid line is the same
in all of the  panels (initial $\feh\ = -1.9;\, \alpha = 1.75;\, 13\,$Gyr).
Cool main sequence stars are on
the extreme right of this diagram, while red giant branch stars are
located slightly above the main sequence stars for $\mathrm{T_{eff}}
\la 5800\,$K.
 \label{teff}}
\end{figure}

The principal results of this paper are shown in Figures \ref{logg} --
\ref{teff} which show the change in the surface iron abundance as a
function of $\log g$ (Figure \ref{logg}) and the effective temperature (Figure
\ref{teff}).  Since stars on the giant branch and on the main sequence
can have the same effective temperatures, it is easiest to illustrate
the effects of heavy element diffusion on the surface iron abundance
when the surface \feh\ value is plotted as a function of $\log g$
(Figure \ref{logg}).  The lowest mass stars (which are on the main
sequence) have the highest surface gravities.  As one goes to lower
surface gravities (to the right in the figure), the stars are more and
more massive.  Low mass main sequence stars have  fairly large 
convection zones.  As a result, the time scale for the diffusion of
metals out of the surface convection zone is very long resulting in
little change in the surface abundance of iron.  The higher mass stars
on the main sequence have less massive surface convection zone,
leading to shorter diffusion time scales and a larger depletion of iron
at the surface.  The maximum depletion in the surface \feh\ abundance
occurs around the main sequence turn-off, where the depth of the
convection zone is at a minimum.  As stars evolve past the turn-off
point, their convection zone rapidly deepens, dredging up the iron
which had diffused out of the surface convection zone on the main
sequence.  As a result, the surface iron abundance returns to its
initial value on the giant branch ($\log g \la 3.6$).

The top panel in Figure \ref{logg} illustrates how the depletion of
\feh\ varies as a function of the assumed age of NGC 6397.  If NGC
6397 is 11 Gyr old (as suggested by the isochrone fits), then the
 turn-off stars have a mass of $M \simeq 0.80\,\mathrm{M}_\sun$
and during their main sequence lifetime have an average convective envelope
mass of around $M_{\mathrm{CZ}} \sim  1\times 10^{-3}\,\mathrm{M}_\sun$
leading to large depletions in \feh\ in the turn-off stars.  In contrast, 
13 Gyr turn-off stars have a mass of $M \simeq 0.76\,\mathrm{M}_\sun$
and during their main sequence lifetime have a convective envelope
mass which is about 3 times as massive as a $M \simeq
0.80\,\mathrm{M}_\sun$ star. As a consequence, if NGC 6397 is really
13 Gyr old, then the predicted \feh\ depletion at the surface is
substantially smaller than if it is 11 Gyr old.  In general, older
ages for globular clusters imply that the effects of atomic diffusion
are smaller.

The middle panel in Figure \ref{logg} illustrates how the depletion of
\feh\ varies as a function of the mixing length assumed in the models.
Lower values of the mixing length lead to less massive convection
zones on the main sequence.  As a result, the time scale for diffusion
is shorter in models with lower mixing lengths, leading to larger
depletions of the surface iron abundance.  Finally, the bottom panel
in Figure \ref{logg} illustrates the effect that changing the initial
\feh\ abundance has on the predicted \feh\ depletions.  As is well
known, the higher the heavy element abundance, the higher the opacities are 
\citep[e.g.\ ][]{opal1}
leading to more massive convection zones and less depletion in the
surface iron abundance.

The effective temperature of a star can be determined from
observations much more accurately than its surface gravity.  As a
result, our model calculations are best compared to observations using
temperature as the independent variable.  This is shown in Figure
\ref{teff} which plots the change in the surface \feh\ abundance as a
function of $\mathrm{T_{eff}}$.  Of course, stars on the giant branch
and the main sequence may have the same temperatures, and so the
depletion lines are  double value functions.  The lowest mass models
on the main sequence are located at the lowest effective temperatures
(extreme right) in Figure \ref{teff}.  It turns out that even the
lowest mass models we evolved had less massive convection zones than
the models on the giant branch. As a consequence, the surface iron
abundance is somewhat higher ($\sim 0.04\,$dex) in the cool giant
branch models ($\mathrm{T_{eff}} \la 5800\,$K) than in the main
sequence models.  Note that 
the difference in surface \feh\ between turn-off stars and giant
branch stars in the models plotted in Figure \ref{teff} is always
greater than $0.3\,$dex.

\cite{gratton} determined the temperature of the main sequence
turn-off stars to be $\mathrm{T_{eff}} = 6476\pm 90\,$K, while 
the giant branch stars they observed had $\mathrm{T_{eff}} = 5478\pm
60\,$K.    For the iron abundances \cite{gratton} found that the
turn-off stars had $\feh = -2.02\pm 0.01$, while the giant branch
stars had $\feh = -2.05\pm 0.03$.  The quoted errors are
internal, given by the standard deviation in the mean for each group
of stars.  In terms of possible systematic errors between the two
\feh\ measurements, this is likely to be dominated by uncertainties in the
adopted temperatures.  \cite{gratton} show that for the turn-off
stars, their error of $\pm 90\,$K in the temperature translates into
an error of $\pm 0.09\,$dex.  For the giant branch stars, their error
of $\pm 60\,$K in temperature translates into an error of $\pm
0.06\,$dex.  Adding these errors in quadrature, one can conclude that
the difference in the iron abundance between the giant
branch and turn-off stars is $\Delta \feh  = 0.03\pm 0.11\,$dex.

\begin{deluxetable}{ccrc}
\tablecaption{Difference in \feh\ between 
turn-off and giant branch models \label{results}}
\tablewidth{0pt}
\tablehead{
\colhead{Initial \feh} & \colhead{Age (Gyr)}   & \colhead{$\alpha$}   &
\colhead{$\Delta \feh$} 
}
\startdata
$-1.9$ &   11 & 1.50 & 1.58\\
$-1.9$ &   11 & 1.75 & 0.92\\
$-1.9$ &   13 & 1.75 & 0.50\\
$-1.9$ &   13 & 2.00 & 0.41\\
$-1.9$ &   15 & 2.00 & 0.31\\
$-2.1$ &   11 & 1.50 & 2.56\\
$-2.1$ &   13 & 1.50 & 0.79\\
$-2.1$ &   13 & 1.75 & 0.57\\
$-2.1$ &   13 & 2.00 & 0.45\\
$-2.1$ &   15 & 2.00 & 0.33\\

 \enddata

\end{deluxetable}

\begin{figure}[t]
\plotone{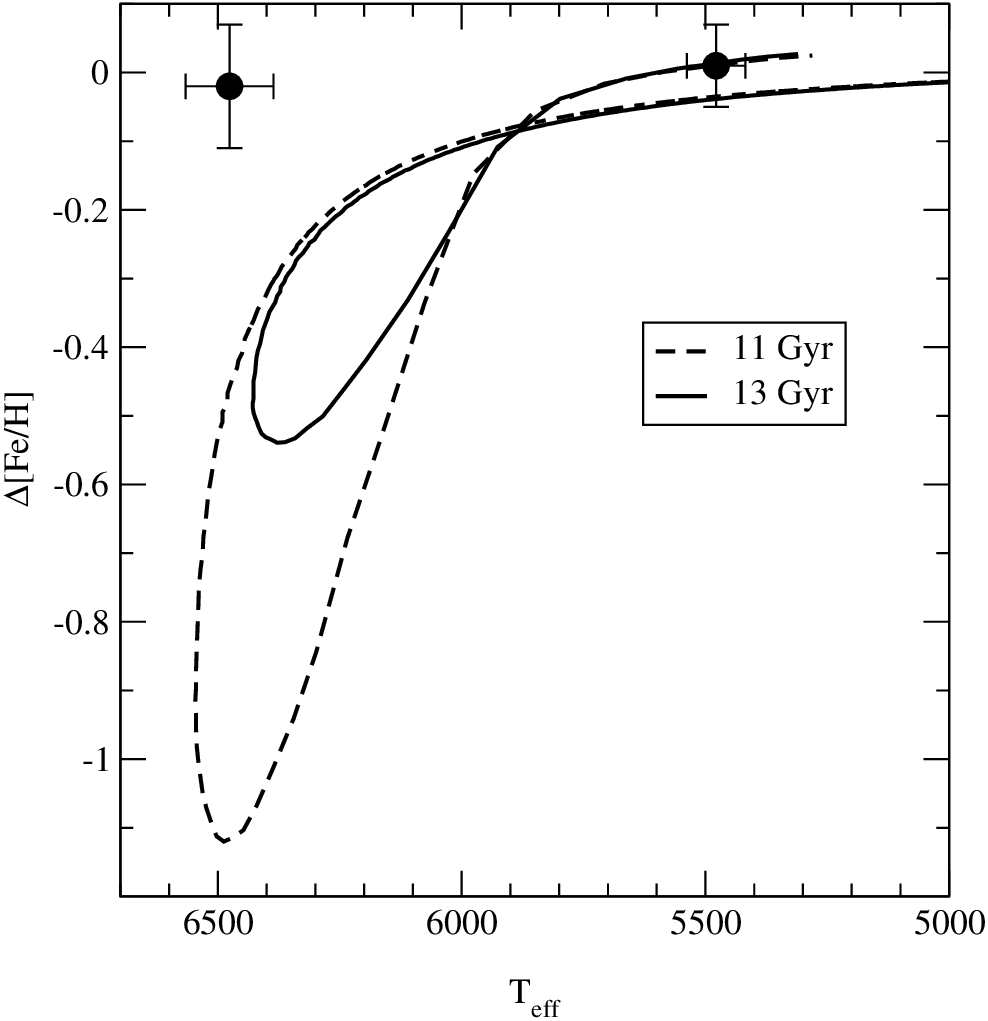}
\caption{The change in the surface \feh\ in the models (solid line
with initial $\feh\ = -1.9$, mixing length $\alpha = 1.75$ and an age
of $13\,$Gyr) is compared to the observations of turn-off and
sub-giant branch stars in the globular cluster NGC 6397
\protect\citep{gratton}.  The models clearly do not match the observed
data.  \label{match}}
\end{figure}

In order to obtain a best estimate for the predicted $\Delta \feh$
between the giant branch and turn-off stars, the models with initial
iron abundances of $\feh = -1.9$ and $-2.1$ were searched to determine
which cases had turn-off stars in the $\pm 1\,\sigma$ range given by
\cite{gratton} ($\mathrm{T_{eff}} = 6386 - 6556$). For each set of
models, $\Delta\feh$ was determined by taking the difference in \feh\
between the turn-off stars and those stars on the giant branch which
were $1000$ K cooler than the turn-off stars.  The results are shown
in Table \ref{results}.  The predicted $\Delta \feh$ values ranged
from 0.31 dex to 2.56 dex, with an average of 0.77 dex.  The minimum predicted
change in \feh\ between the giant branch and the turn-off stars is
$2.5\,\sigma$ larger than the observed difference.  The lowest $\Delta
\feh$ values occur for the 15 Gyr models; however as discussed in
the previous section the 15 Gyr isochrones do not fit the observed
color-magnitude diagram. For models 13 Gyr or younger, the minimum
predicted $\Delta\feh = 0.41\,$dex, which is $3.5\,\sigma$ larger than
observed.  The difficulty in matching the observed \feh\ values is
shown in Figure \ref{match} which compares models which had an initial
\feh\ of $-1.9$, age of 13 Gyr and a mixing length of $\alpha = -1.75$
to the data from \cite{gratton}.

An examination of Figure \ref{logg} shows that the minimum possible
depletion will occur in the 15 Gyr, $\alpha = 2.00$ models which had
an initial iron abundance of $\feh = -1.7$. These models have a
turn-off temperature of $\mathrm{T_{eff}} = 6361\,$K, which differs by
$1.3\,\sigma$ from the observed value, and a giant branch \feh\ which
differs from the observed value by $3\,\sigma$.  In this extreme case,
the predicted change in \feh\ between the turn-off and giant branch is
$\Delta \feh = 0.28\,$dex, which is $2.3\,\sigma$ different from the
observed value.  Since all of the models differ by more than
$2\,\sigma$ from the observations, we conclude that heavy element
diffusion does not occur near the surface of metal-poor stars.

\section{Discussion \label{secage}}
Observations of Li in metal-poor field stars have suggested for some
time that diffusion is inhibited near the surface of metal-poor stars
\citep[e.g.\ ][]{Michaud,con,chab94}.  However, \cite{sw2001} have
pointed out that the when one takes into account the observational
errors, uncertainties in the Li abundance determinations and in the
effective temperature scale, and the size of the observed samples
of stars, the Li abundance observations can be reproduced
by models which include fully efficient diffusion. 
In contrast, observations of iron abundances in turn-off and giant branch stars
in the metal-poor globular cluster NGC 6397 clearly show that  
heavy element diffusion does not occur near the surface of metal-poor
stars.  

A key difference between the Li observations analyzed by
\cite{sw2001} and the Fe observations analyzed here is that the Li
observations are primarily obtained for field stars, while the Fe
abundances which have been analyzed in this paper have been obtained
for a single globular cluster.  Evolutionary time scales near the main
sequence turn-off are much more rapid than on the main sequence.  As a
result observations of random field stars are biased against finding
stars near the turn-off.  It is stars at the turn-off in which
diffusion causes the greatest depletion of Li.  Hence, observations of
field stars are biased against finding Li depleted stars.  The Monte
Carlo simulations of \cite{sw2001} show that even if Li diffusion
occurs at the surface of metal-poor stars, observational selection
effects imply that one is unlikely to observe stars with large Li
depletions, given the size of present day samples.  \cite{sw2001}
found that three times as many stars as are in the current datasets
must be observed in order for the observational data to show clear
signs of Li depletion at hot temperatures. In contrast,
\cite{gratton} were able to use the observed color-magnitude diagram
of NGC 6397 to select turn-off stars for spectroscopic study.  Hence,
these observations are a much cleaner test for the presence of
diffusion than the field star Li abundance measurements analyzed by
\cite{sw2001}.  

Iron is not subject to depletion on the red giant branch making it
easy to compare the observed iron abundances between the main sequence
turn-off and giant branch in a given cluster.  The
deepening of the convection zone on the red giant branch leads to
significant Li depletion in these stars.  The globular cluster
Li abundance data analyzed by \cite{sw2001} involved a comparison
between giant branch stars and main sequence turn-off stars.  As the
giant branch stars undergo  significant Li depletion these
observations do not serve as strong
constraints on the presence of diffusion.  As a result, there is no
conflict between the study of \cite{sw2001} and this study.
\cite{sw2001} simply claim that the available Li data do not rule out
the presence of diffusion near the surface of metal-poor stars.  The
iron abundances measurements analyzed in this paper are much more
sensitive to the presence of diffusion than the Li observations.
Hence, the present observational data shows that diffusion is
inhibited near the surface of metal-poor giant branch stars.

Two mechanisms which inhibit diffusion in the surface
layers of stars are modest amounts of mass loss \citep{wind} and
mixing induced by rotation \citep[e.g.\ ][]{rotv,pin1,pin2}.  The wind
loss models require that the mass loss in low mass metal-poor stars be
an order of magnitude larger than is observed in the Sun.  Observations
of a correlated depletion of Lithium and Beryllium in Population I F
stars suggest that rotationally induced mixing, and not mass loss is
operating in these stars \citep{con2}.  

Mixing induced by rotation suppresses diffusion by effectively
increasing the mass of the surface mixed zone, leading to longer time
scales for the depletion of elements at the surface of the star (cf.\
equation \ref{eqtau}).  Using equation (\ref{eqtau}) as a guide, we
may estimate the mass of the surface mixed zone in metal-poor stars by
requiring that the surface abundance of iron change by less than 0.1
dex in 13 Gyr. The constant $K$ in equation (\ref{eqtau}) was
calculated for our models, and from this, we estimate that the mixed
zone at the surface of a metal-poor star has a mass $M_{mix} \ga
0.005\,M_\sun$.  In contrast, the convection zone masses in turn-off
stars with $\feh \simeq -2.0$ are $0.002\,M_\sun$ or smaller.  An
upper limit to the mass of the mixed zone may be determined from solar
observations.  Helioseismology clearly shows that diffusion does occur
in the Sun \citep{jcd,basu2}.  The mass of the convection zone in the
Sun is $M_{CZ} =
0.02\,M_\sun$.  As diffusion is not inhibited in the Sun, this
suggests that the mechanism which inhibits diffusion only acts in the
very outermost layers of a star (ie: for $M > 0.98\,M_\sun$).
Combined with the estimate for the minimum mass of the mixed zone
given above, it appears that the mass of the mixed zone at the surface
of hot metal-poor stars is in the range $M_{mix} \simeq 0.005 -
0.02\,M_\sun$.

\begin{figure}[t]
\plotone{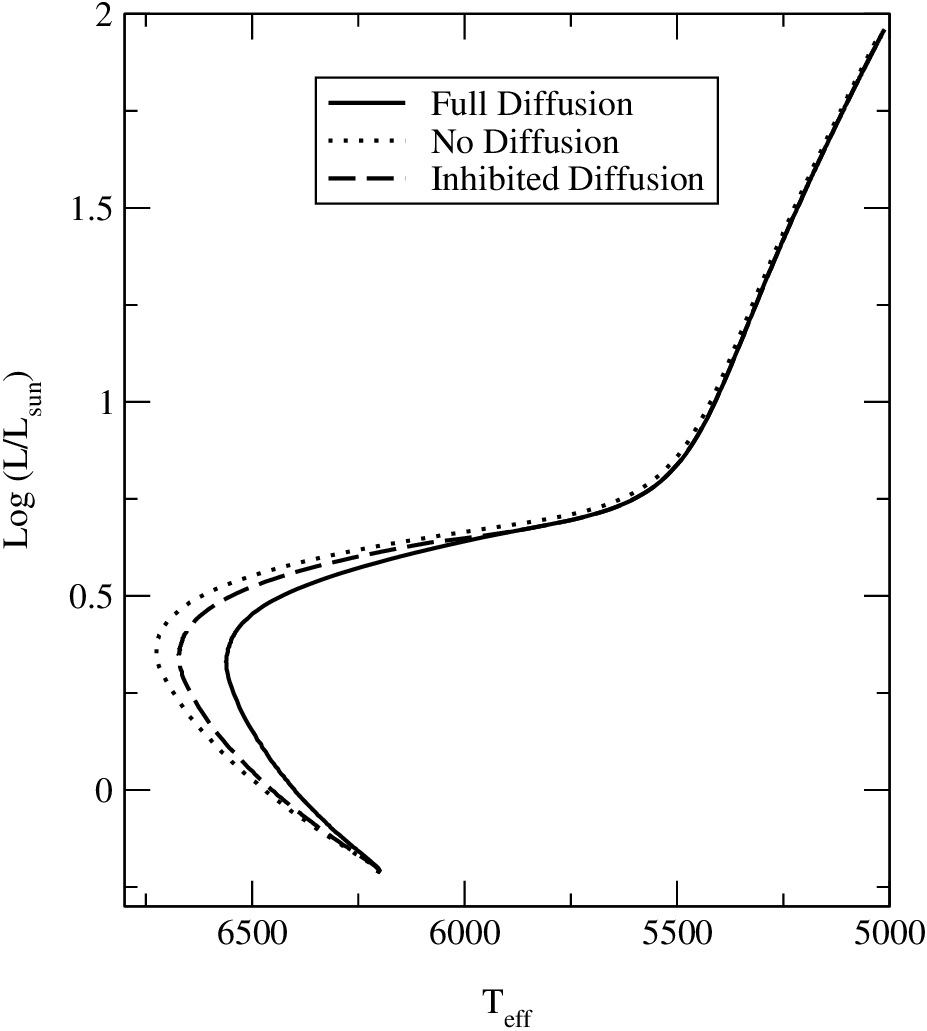}
\caption{
The evolution of models with  full diffusion, no diffusion, and
diffusion which has been inhibited in the outer layers of the star.
All models had a mass of $M = 0.80\,M_\sun$, an initial iron abundance
of $\feh = -1.9$ and a mixing length of $\alpha = 1.75$.  
 \label{nodiff}}
\end{figure}

To determine what effects inhibiting diffusion in the outer layers of
a star has on the observed properties of the models, the diffusion
subroutine in our code was altered so that diffusion was shut off in
the outer layers of the star.  This was accomplished by setting the
diffusion coefficients to zero when $M > M_* - M_{mix}$, where $M_*$
is the total mass of the star.  The diffusion coefficients were ramped
from zero, up to their standard value in the region $M_* - M_{mix} > M
> M_* - M_{ramp}$.  For these test runs, we used $M_{mix} =
0.005\,M_\sun$ and $M_{ramp} = 0.01\,M_\sun$.  The results of a sample
calculation are shown in Figure \ref{nodiff}, which shows the evolution
in the effective temperature-luminosity plane.  We see that when
diffusion is inhibited in the outer layers of the model, the predicted
temperatures of the models are similar to models evolved without
diffusion.  The primary reason diffusion models are cooler
than non-diffusion models is that diffusion of helium out of the
envelope increases the envelope opacity, which increases the model
radius and hence, decreases the effective temperature.  Thus, when
diffusion is inhibited in the outer layers, the effective temperature
of the models resembles the non-diffusion models.

The model in which diffusion was inhibited at the surface had a 
similar lifetime to the full diffusion model.  This is
because diffusion was not inhibited in the deep interior of the model.
Helium diffused into the core (displacing the hydrogen) which leads
to a shorter main sequence lifetime compared to non-diffusion models.
Such tracks by themselves, however, may provide a misleading guide to
the impact of diffusion on cluster age estimates. To determine
globular cluster ages one compares isochrones to observations.
Diffusion will change the shape of the isochrones in a different way
than it changes the shape of individual tracks because models with
different masses experience different degrees of diffusion.

\begin{figure}[t]
\plotone{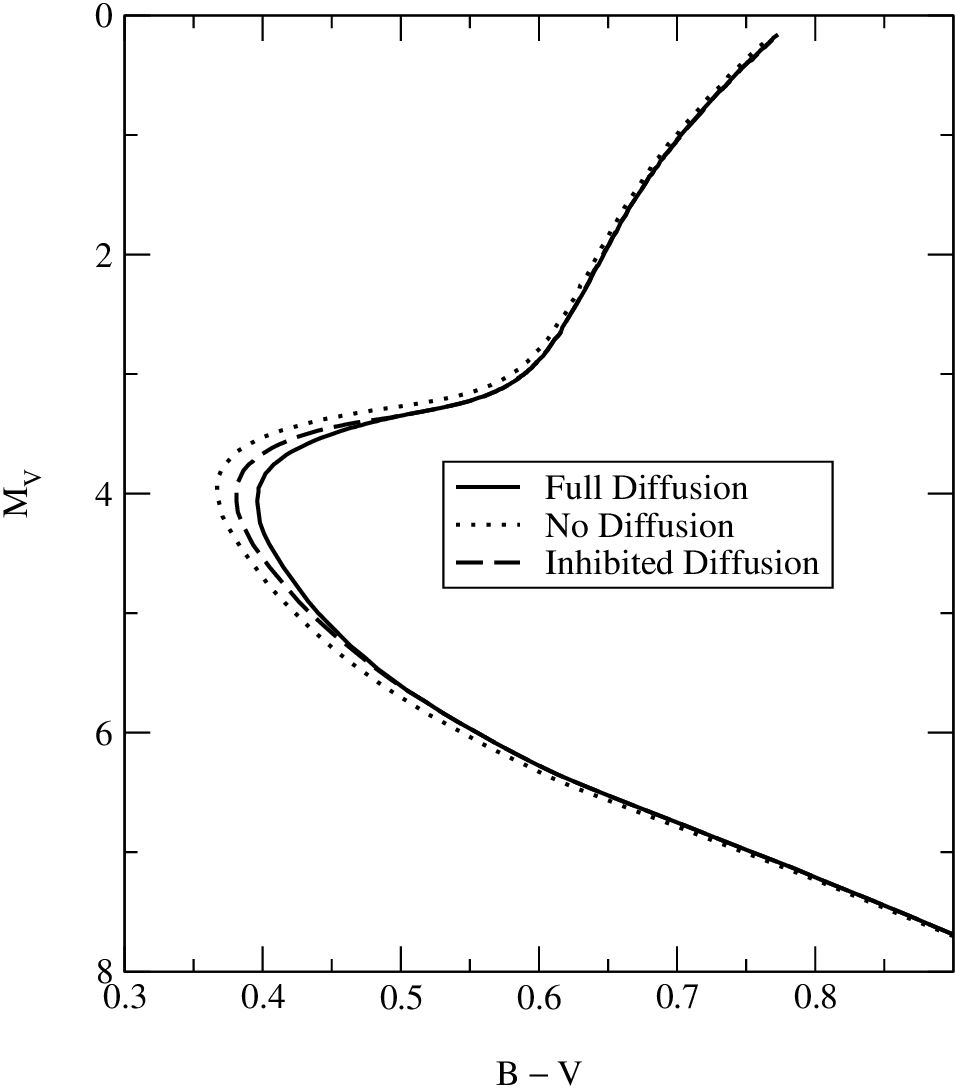}
\caption{Isochrones with an age of 13 Gyr, an initial iron abundance
of $\feh = -1.9$ and a mixing length of $\alpha = 1.75$.
 \label{isod_nod}}
\end{figure}

To explore the consequences this modification of diffusion has on age
determinations, isochrones were calculated based upon the no diffusion
models, and models in which diffusion was inhibited in the outer
layers.  These isochrones are compared to the full diffusion
isochrones in Figure \ref{isod_nod}.  The isochrone in which diffusion
has been inhibited falls half-way between the isochrone with full
diffusion, and the isochrone with no diffusion.  If \mvto\ is used as
an age indicator, then ages determined using isochrones in which
diffusion has been inhibited will be 4\% larger ages derived from the
full diffusion isochrones, and 4\% smaller than ages derived using the
no diffusion isochrones.

If diffusion is inhibited near the surface of metal-poor stars
(likely by rotation induced mixing), one naturally wonders if
diffusion occurs  in the deep interior of metal-poor stars.  Perhaps the
rotation induced mixing suppress the diffusion throughout the entire
star.  At the present time, there is no clear answer as to whether or
not diffusion is occurring in the deep interior of metal-poor stars.  A
definitive answer will likely have to await observations from stellar
seismology,  which will probe the interior structure of metal-poor stars
and allow one to test the interior properties of theoretical models.

This study has shown that diffusion is inhibited in the surface layers
of metal-poor stars.  Stellar models and isochrones which include the
full effects of diffusion are in error, and should not be used for
comparing to observational data.  Our calculations in which diffusion
is inhibited near the surface of the star are admittingly ad-hoc.
However, as helioseismology has shown that diffusion is occurring in
the interior of the Sun, we believe it is likely that diffusion is
also occurring in the interiors of metal-poor stars.  The obvious
difference between the Sun and the metal-poor turn-off stars, is that
the surface convection zone is much larger in the Sun.  Theoretical
calculations of rotation induced mixing have shown that this mixing is
most effective in the outer layers of a star \citep[e.g.\
][]{chab95,zahn}.  The fact that diffusion is occurring in the interior
of the Sun makes us believe that diffusion is only being inhibited in
the outermost layers of stars.  For this reason, we prefer the use of
stellar models and isochrones in which diffusion operates in the
interior of the model (but not in the outer layers) to models which do
not include diffusion.  Given that diffusion is occurring in the Sun,
stellar models and isochrones which include diffusion are 
appropriate for solar-type stars.  The model we have presented for
inhibiting diffusion in the surface layers of stars allows one to use
a set of isochrones calculated using the same assumptions for both
metal-poor and solar-type stars.

\section{Summary \label{secsumm}}
A large number of stellar evolution models which include the effects
of helium and heavy element diffusion were constructed.  The amount of
diffusion of iron out of the surface convection zone is a strong
function of the mass and age of a star.  As a result, the models
predicted that the \feh\ abundance measured in the turn-off stars in
the globular cluster NGC 6397 should be more than 0.28 dex lower than
the \feh\ measured in the giant branch stars.  In contrast,
observations by \cite{gratton} demonstrate that the turn-off and giant
branch stars have identical values of \feh\, indicating that
uninhibited metal diffusion does not occur in the surface layers of
metal-poor stars.  Based upon this observed fact and the physical
principals underlying diffusion processes, we estimate that the
minimum mass of the region where diffusion is inhibited near the
surface of metal-poor stars to be $M_{mix} \ga 0.005\,M_\sun$.  Models
and isochrones in which diffusion was inhibited in the outer layers
were calculated and compared to full diffusion isochrones, and
isochrones without diffusion.  The isochrones in which diffusion has
been inhibited have properties which lie half-way between the full
diffusion isochrones, and the no diffusion isochrones.  As a
consequence, globular cluster age estimates which use \mvto\ 
as their age indicator, and which are based upon
models which include the full effects of diffusion need to be revised
upward by 4\%.  Helioseismic observations clearly show that diffusion
is occurring in the Sun, which suggests that diffusion operates in
most regions of a star.  Our models which include
diffusion which is inhibited in the outer layers of the star yield
globular cluster ages which are 4\% smaller than models which do not
include diffusion when \mvto\ is used as an age indicator.

\acknowledgments
We would  like to thank the  anonymous referee for  helpful
comments on an earlier version of this manuscript.
Research supported in part by a NSF CAREER grant 0094231  to Brian Chaboyer.  
Dr.\ Brian Chaboyer is a Cottrell Scholar of the Research Corporation.  



\clearpage 

\end{document}